\documentclass[%
 reprint,
nofootinbib,
 amsmath,amssymb,
 aps,
]{revtex4-2}

\usepackage{xcolor}
\usepackage{graphicx}
\usepackage{dcolumn}
\usepackage{bm}
\usepackage{booktabs}
\usepackage{hyperref}

\usepackage{todonotes}

\begin{document}

\title{Neutrinos in Cosmology}

\author{Eleonora Di Valentino}
\email{e.divalentino@sheffield.ac.uk}
\affiliation{School of Mathematics and Statistics, University of Sheffield, Hounsfield Road, Sheffield S3 7RH, United Kingdom}

\author{Stefano Gariazzo}
\email{stefano.gariazzo@ift.csic.es}
\affiliation{%
Instituto de Fisica Teorica, CSIC-UAM
C/ Nicolás Cabrera 13-15, Campus de Cantoblanco UAM, 28049 Madrid, Spain
}%
\author{Olga Mena}
\email{omena@ific.uv.es}
\affiliation{Instituto de F{\'\i}sica Corpuscular  (CSIC-Universitat de Val{\`e}ncia), E-46980 Paterna, Spain}%

\date{\today}

\begin{abstract}
Neutrinos are the least known particle in the Standard Model of elementary particle physics. They play a crucial role in cosmology, governing the universe's evolution and shaping the large-scale structures we observe today. In this chapter, we review crucial topics in neutrino cosmology, such as the neutrino decoupling process in the very early universe. We shall also revisit the current constraints on the number of effective relativistic degrees of freedom and the departures from its standard expectation of 3. Neutrino masses represent the very first departure from the Standard Model of elementary particle physics and may imply the existence of new unexplored mass generation mechanisms. Cosmology provides the tightest bound on the sum of neutrino masses, and we shall carefully present the nature of these constraints, both on the total mass of the neutrinos and on their precise spectrum. The ordering of the neutrino masses plays a major role in the design of future neutrino mass searches from laboratory experiments, such as neutrinoless double beta decay probes. Finally, we shall also present the futuristic perspectives for an eventual direct detection of cosmic, relic neutrinos.
\end{abstract}

\newcommand{\cnb}{C$\nu$B}
\newcommand{\Neff}{\ensuremath{N_{\rm eff}}}
\newcommand{\Neffstd}{\ensuremath{N_{\rm eff,std}}}
\maketitle


\section{Neutrino decoupling in the early universe}
\label{sec:nu_decoupling}
Neutrinos have contributed to the evolution of the universe since its earliest times. They play an important role in several processes, and their presence leaves a characteristic imprint on several observables.

When the temperature of photons was above a few MeV during radiation domination, neutrinos were coupled to the electromagnetic plasma due to their weak interactions with electrons and positrons. When these interactions fell below the expansion rate of the universe, neutrinos decoupled from the thermal plasma at a temperature around 2~MeV and started to propagate freely until today.
These neutrinos constitute the Cosmic Neutrino Background (\cnb), which we will discuss in more detail later on. As neutrino decoupling nears its end, electrons and positrons start to become non-relativistic, transferring their energy density to photons while annihilating away. Since neutrinos with the highest momenta are still interacting with the thermal plasma at this stage, they receive a small fraction of this entropy, causing their momentum distribution function to be slightly distorted from the equilibrium Fermi-Dirac distribution.

After these two processes are complete, Big Bang Nucleosynthesis (BBN) occurs until the photon temperature drops below approximately $0.05$ MeV, leading to the production of light nuclei. Finally, after the end of radiation domination (at $T \sim 1$ eV), the last scattering of photons occurs ($T \sim 0.3$ eV), and the Cosmic Microwave Background (CMB) radiation is produced. The presence of neutrinos affects all of the above-mentioned processes through their contribution to the total radiation energy density, which controls the expansion rate of the universe during radiation domination.
Observations of BBN abundances or the CMB spectrum, therefore, can provide us with information about the neutrino contribution to early universe physics.

The last two particles that remain relativistic after electrons and positrons disappear are neutrinos and photons. In case other relativistic particles exist, as we will discuss in the following, they are normally categorized under the name "dark radiation," since they do not take part in electroweak interactions.

The amount of the radiation energy density, $\rho_{\rm R}$, is commonly parameterized in terms of the effective number of relativistic degrees of freedom, \Neff, by
\begin{equation}
\rho_{\rm R} = \rho_\gamma \left( 1 + \frac{7}{8} \left(\frac{4}{11}\right)^{4/3} \Neff \right)\,,
\end{equation}
where $\rho_\gamma$ represents the photon energy density. Here, \Neff\ includes the contribution of all relativistic particles besides photons. If we consider the simplest three-neutrino case, with an instantaneous decoupling process, \Neff\ would be equal to 3. A value $\Neff \neq 3$, however, would be the consequence of either new degrees of freedom which have nothing to do with standard neutrinos, or a non-standard momentum distribution for the three neutrinos.

Even in the case with three neutrinos, the value of \Neff\ deviates from 3 because of non-instantaneous decoupling. The deviation can be computed numerically by taking into account the full framework of neutrino oscillations, interactions with electrons and positrons, Finite-Temperature corrections to Quantum Electro-Dynamics (FT-QED), and the expansion of the universe. The momentum-dependent calculation is numerically challenging but can be solved to a very high level of precision. The standard value for \Neff\ is computed to be $\Neffstd = 3.044$~\cite{Akita:2020szl, Froustey:2020mcq, Bennett:2020zkv}, see also~\cite{Cielo:2023bqp, Drewes:2024wbw}. This number was previously claimed to be a bit higher~\cite{Mangano:2005cc, deSalas:2016ztq}, but state-of-the-art calculations confirm that the (theoretical and numerical) error on \Neffstd\ is at the level of $10^{-4}$, including variations of the fundamental constants of physics and the effect of neutrino oscillation parameters within the currently allowed observational range by terrestrial experiments.
Even in the presence of three neutrinos, this number can vary due to non-standard interactions (NSI, see e.g.,~\cite{Farzan:2017xzy}) between neutrinos and electrons~\cite{Du:2021idh, deSalas:2021aeh}. The effects of NSI on \Neff\ are rather small and do not significantly impact the $H_0$ value through its correlation with \Neff, see also the incoming section for more details. The precision that next-generation CMB measurements~\cite{Ade:2018sbj, CMB-S4:2016ple} will achieve, however, will be sufficient to test some of these scenarios. Also, the presence of additional neutrino states alters the standard value of \Neff, see e.g.,~\cite{Gariazzo:2019gyi, Gariazzo:2022evs}. We will discuss constraints on additional neutrino species in the upcoming section.

After they decouple as relativistic particles, their momentum distribution function maintains the same shape until today, while the temperature of relic neutrinos decreases according to the expansion of the universe. Nowadays, \cnb\ neutrinos are expected to have a temperature of approximately 1.9~K or $10^{-4}$~eV, which is smaller than the temperature of CMB photons by a factor of $\sim1.4$. Given the small temperature and the existence of neutrino oscillations (see e.g.,~\cite{Esteban:2020cvm, deSalas:2020pgw, Capozzi:2021fjo, Gonzalez-Garcia:2021dve}), which implies a mass above at least $\sim8$~meV and 50~meV for the second and third neutrino mass eigenstates respectively, at least two out of three \cnb\ neutrinos are non-relativistic today. This means that they must have undergone a non-relativistic transition at some point during the matter domination epoch, thus leaving an imprint on the growth of structures, as we will discuss in the following.

If we use their temperature to compute the neutrino number density in empty space, we obtain that the \cnb\ neutrinos are the second most abundant particles in the Universe, with a number density of 56~cm$^{-3}$ per degree of freedom. Since non-relativistic neutrinos feel the gravitational attraction of local structures, local overdensities can grow with respect to the average value of their number density. The overdensity depends on the mass of each neutrino and on the total gravitating mass in the local object. The neutrino clustering in the neighborhoods of Earth has been computed by means of $N$-``one-body'' simulations, see e.g.,~\cite{Ringwald:2004np, deSalas:2017wtt, Zhang:2017ljh, Mertsch:2019qjv}. Early studies considered a simplified spherically symmetric scenario with the Milky Way as the only local source of gravitational attraction~\cite{Ringwald:2004np, deSalas:2017wtt, Zhang:2017ljh}, while more recently a back-tracking method allowed for a study that also takes into account the effect of the Virgo cluster and the Andromeda galaxy~\cite{Mertsch:2019qjv}, which however have been proven to be secondary with respect to the contribution of the Milky Way.
Although in the past, large overdensities were claimed to be possible by incorrectly interpreting the results of~\cite{Ringwald:2004np}, in more recent times it has been clarified that for values of the neutrino mass allowed by terrestrial experiments (see e.g.,~\cite{KATRIN:2021uub}), the increase with respect to the number density in vacuum cannot exceed a factor of a few units. For neutrino masses allowed by cosmological constraints (see section~\ref{sec:mnu_bounds}), instead, the overdensity cannot differ from the empty value by more than $\sim50$\%. Experimental constraints on the local number density of relic neutrinos, however, are very far from the theoretical prediction, see e.g.,~\cite{KATRIN:2022kkv, Tsai:2022jnv}. The value of the local neutrino overdensity is very important when determining the perspectives for the direct detection of relic neutrinos, which we will discuss in section~\ref{sec:direct_detection}.

\section{Bounds on the Number of neutrino species, \Neff}
\label{sec:neff_bounds}

If we consider the presence of additional particles (axions~\cite{OHare:2024nmr,DEramo:2022nvb,Notari:2022ffe,Giare:2020vzo}, sterile neutrinos~\cite{Wang:2024hks,Pan:2023frx,diValentino:2022njd,DiValentino:2021rjj}, and so on) or more complicated neutrino interactions, for example with dark matter~\cite{Brax:2023tvn,Giare:2023qqn,Mosbech:2020ahp,Stadler:2019dii,DiValentino:2017oaw}, one would expect larger contributions to \Neff\ than those related to the decoupling process~\cite{Gariazzo:2023hch}, and therefore they could be testable by current and future cosmological measurements.

Cosmology provides bounds on the relativistic degrees of freedom $N_{\rm eff}$ based on two distinct epochs: the BBN epoch, which occurred during the very first three minutes of the universe's evolution, and the CMB epoch, which took place when the age of the universe was four hundred thousand years old, when electrons and protons combined to form neutral hydrogen for the first time. The effective number of neutrinos also affects the fluctuations of the matter perturbations, albeit in a subdominant manner.

Concerning BBN bounds, they are based on the abundances of the first light nuclei (heavier than the lightest isotope of hydrogen), which were synthesized in the very early universe. The abundances of these BBN elements therefore provide a cosmological laboratory where to test extensions to the minimal $\Lambda$CDM scenario involving, in general, additional relativistic species contributing to \Neff. Indeed, these additional contributions to the dark radiation of our universe will increase the expansion rate $H(z)$ and will anticipate the period of weak decoupling, implying a larger freeze-out temperature of the weak interactions. In turn, this will lead to a higher neutron-to-proton ratio, and consequently to a larger fraction of primordial Helium and Deuterium (as well as to a higher fraction of other primordial elements) with respect to hydrogen. 
This makes BBN a laboratory where to test for additional contributions to \Neff, present in beyond-the-Standard Model physics frameworks: given a concrete model, by means of the resolution of a set of differential equations governing the nuclear interactions in the primordial plasma (see e.g.,~\cite{Pisanti:2007hk, Consiglio:2017pot, Gariazzo:2021iiu}), it is possible to compute the light element abundances and compare the results to the values inferred by astrophysical and cosmological observations. Given current uncertainties, the standard BBN predictions show a good agreement with direct measurements of primordial abundances of both Deuterium and Helium~\cite{Pitrou:2020etk, Mossa:2020gjc, Pisanti:2020efz, Yeh:2020mgl}, limiting $\Delta \Neff \lesssim 0.3-0.4\%$ at 95\% CL. 
BBN predictions for the Helium abundance ($Y_p^{\rm BBN}$) also play a role in the CMB angular spectra, as the baryon energy density can be computed via the simple formula~\cite{Serpico:2004gx}:
\begin{equation}
    \Omega_b h^2 = \frac{1 - 0.007125\ Y_p^{\rm BBN}}{273.279}\left(\frac{T_{\rm CMB}}{2.7255\ \mathrm{K}}\right)^3 \eta_{10} \ ,
    \label{eq.Yp}
\end{equation}
where $\eta_{10} \equiv 10^{10}n_b/n_\gamma $ is the baryon-to-photon ratio today, $T_{\rm CMB} $ is the CMB temperature at the present time, and $Y_p^{\rm BBN} \equiv 4 n_{\rm He}/n_{b}$ is the final Helium \textit{nucleon fraction}, defined as the ratio of the 4-Helium number density to the total baryon one. 

Concerning the CMB temperature power spectrum, first of all, varying \Neff\ changes the redshift of the matter-radiation equivalence, $z_{\rm eq}$, inducing an enhancement of the early Integrated Sachs Wolfe (ISW) effect which increases the CMB spectrum around the first acoustic peak. Namely, in the fully matter-dominated period, the gravitational potentials are almost constant in time and therefore the ISW effect, which is sensitive to the time variation of the gravitational potentials, will be very small. Right after recombination, there is still a radiation component present in the universe, and assuming a vanishing anisotropic stress and evaluating the Bessel function at recombination ($\eta_r$), the early ISW effect leads to a CMB temperature perturbation which reads as:

\begin{equation}
 \Theta_\ell (k)\simeq 2 j_\ell (\eta_r) \left[\Phi (k, \eta_m)-\Phi(k,\eta_r)\right]~,
\end{equation}
where the gravitational potential $\Phi$ is evaluated in the matter-dominated regime, $\eta_m$, and $j_\ell$ refer to the Bessel functions. 
Notice that this early ISW effect adds in phase with the primary anisotropy, increasing the height of the first acoustic peaks, with an emphasis on the first one, due to the fact that the main contribution of the ISW effect is at scales $k\sim 1/\eta_e$, i.e., around the first acoustic peak. In addition, the early ISW effect will be suppressed by the square of the radiation-to-matter ratio $\propto [(1+z_r)/(1+z_{eq})]^2$, i.e., a larger (smaller) matter component will result in a smaller (larger) ISW amplitude due to the larger (smaller) value of $z_r$. The enhancement factor of the ISW effect amplitude when $\Delta \Neff >0$ makes this effect an excellent observable to identify extra relativistic particles present at recombination. Nevertheless, this is a sub-dominant effect in the overall impact of the effective number of relativistic degrees of freedom on the CMB.

Ref.~\cite{Hou:2011ec} provides a detailed explanation concerning the most relevant impact of changing \Neff. Indeed, its main effect is located at high multipoles $\ell$ rather than at the very first peaks, i.e., at the CMB damping tail. If $\Delta \Neff$ increases, the Hubble parameter $H$ during radiation domination will increase as well. This will induce a delay in the matter-radiation equality and will also modify the sound speed and the comoving sound horizon:
\begin{displaymath}
r_{\rm s} = \int_0^{\tau'} d\tau c_{\rm s} (\tau) = \int_0^a \frac{da}{a^2 H} c_{\rm s}(a)~,
\end{displaymath}
proportional to the inverse of the expansion rate, i.e., $r_{\rm s}\propto1/H$. 
Overall, there will be a reduction in the angular scale of the acoustic peaks $\theta_{\rm s}=r_{\rm s}/D_{\rm A}$, where $D_{\rm A}$ is the angular diameter distance, causing a horizontal shift of the peak positions towards higher multipoles. In addition, Silk damping will affect the height of the CMB high multipole region. Baryon-photon decoupling is not an instantaneous process, leading to a diffusion damping of oscillations in the plasma. If decoupling starts at $\tau_{\rm d}$ and ends at $\tau_{\rm ls}$, during $\Delta\tau$ the radiation free streams on scale $\lambda_{\rm d}=\left(\lambda\Delta\tau\right)^{1/2}$ where $\lambda$ is the photon mean free path and $\lambda_{\rm d}$ is shorter than the thickness of the last scattering surface.
As a consequence, temperature fluctuations on scales smaller than $\lambda_{\rm d}$ are damped because on such scales photons can spread freely both from overdensities and underdensities. The overall result is that the damping angular scale $\theta_{\rm d}=r_{\rm d}/D_{\rm A}$ is proportional to the square root of the expansion rate $\theta_{\rm d}\propto\sqrt{H}$ and consequently it increases with $\Delta \Neff$, inducing a suppression of the peaks located at high multipoles and a smearing of the oscillations that intensifies at the CMB damping tail.

The three aforementioned effects caused by a non-zero $\Delta \Neff$ (namely, the redshift of equivalence, the size of the sound horizon at recombination, and the damping tail suppression) can be easily compensated by varying other cosmological parameters, including the Hubble constant $H_0$~\cite{Gariazzo:2023hch}. Notice that the horizontal shift towards smaller angular scales caused by an increased value of \Neff\ can be compensated by decreasing $D_{\rm A}$, which can be automatically satisfied by increasing $H_0$. The effect of \Neff\ on the damping tail is, however, more difficult to mimic via $H_0$, as it is mostly degenerate with the helium fraction which enters directly in $r_{\rm d}$, i.e., the mean square diffusion distance at recombination via $n_e$, the number density of free electrons.

Nevertheless, there is however one effect induced by \Neff\ which cannot be mimicked by other cosmological parameters: the neutrino anisotropic stress~\cite{Bashinsky:2003tk, Hannestad:2004qu}, related to the fact that neutrinos are free-streaming particles propagating at the speed of light, faster than the sound speed in the photon fluid. This leads to a suppression of the oscillation amplitude of CMB modes that entered the horizon in the radiation epoch. The effect on the CMB power spectrum is therefore located at scales that cross the horizon before the matter-radiation equivalence, resulting in an increase in power of $5/(1+\frac{4}{15}f_\nu)$~\cite{Hu:1995en}, where $f_\nu$ is the fraction of radiation density contributed by free-streaming particles.

All in all, our current knowledge confirms that \Neff\ is close to 3 as measured
by CMB observations ($\Neff=2.99^{+0.34}_{-0.33}$ at 95\% confidence level (CL)~\cite{Planck:2018vyg})
or BBN abundances (e.g., $\Neff=2.87^{+0.24}_{-0.21}$ at 68\% CL~\cite{Consiglio:2017pot}) independently. Furthermore, the above constraints have been shown to be extremely robust against different fiducial cosmologies. 
Ref.~\cite{diValentino:2022njd} reported very similar constraints on extended cosmologies (see Tab.~\ref{tab:tab1}), adapted from the very same reference. In Tab.~\ref{tab:tab1}, $\Omega_k$ refers to the curvature component in the universe, $\alpha_s$ to a possible running of the scalar spectral index, $m_{\nu,s}^{\rm eff}$ to the mass of a sterile neutrino state, $Y_p^{\rm BBN}$ to the BBN primordial Helium fraction, 
$w_0$ to the dark energy equation of state, and $w_a$ to a possible time-variation of the former, i.e., $w(a)=w_0 +w_a (1-a)$. The different possible neutrino mass eigenstate spectra are represented by DH (degenerate spectrum), NH (normal mass ordering, where the lightest mass eigenstate is $m_1$ and the atmospheric mass splitting is positive), and IH (inverted mass ordering, where the lightest mass eigenstate is $m_3$ and the atmospheric mass splitting is negative). As can be noticed, the largest departure concerning the uncertainties in \Neff\ appears in models that consider also the Helium fraction to be a free parameter, due to the degeneracy among \Neff\ and $Y_p^{\rm BBN}$ previously discussed when describing the Silk damping effect.  

\begin{table}
\centering
\begin{tabular}{ccc}
\toprule
\textbf{Cosmological model} &  & \boldmath{$\Neff$}\\
\hline\hline

$+\Neff$ &  & $3.08\pm0.17$ \\[2ex]

$+\Neff+ \sum m_\nu$
&  DH & $3.06\pm0.17$\\
&  NH &$3.11\pm0.17$ \\
&  IH & $3.15\pm0.17$ \\[2ex]

$+\Neff+\Omega_{k}$ &  & $3.04\pm0.19$ \\[2ex]

$+\Neff+\alpha_s$ & & $3.03\pm0.19$  \\[2ex]

$+\Neff+m^{\rm eff}_{\nu, s}$ & & $<3.41$ \\[2ex]

$+\Neff+Y_p^{\rm BBN}$ & & $3.17^{+0.27}_{-0.31}$ \\[2ex]

$+\Neff+w_0$ & & $2.99\pm0.18$  \\[2ex]

$+\Neff+(w_0>-1)$ & &  $3.12\pm0.17$  \\[2ex]

$+\Neff+w_0+w_a$ & & $2.91\pm0.18$ \\[2ex]

\hline
model marginalized &  & $3.07^{+0.19}_{-0.18}$ \\

\bottomrule
    \end{tabular}
    \caption{Constraints at 68\% and upper limits at 95\% CL,
    for the $\Lambda$CDM model plus $\Neff$ model and its extensions (adapted from Ref.~\cite{diValentino:2022njd}).}
\label{tab:tab1}
\end{table}

\begin{figure*}
	\centering
	\includegraphics[scale=0.26]
{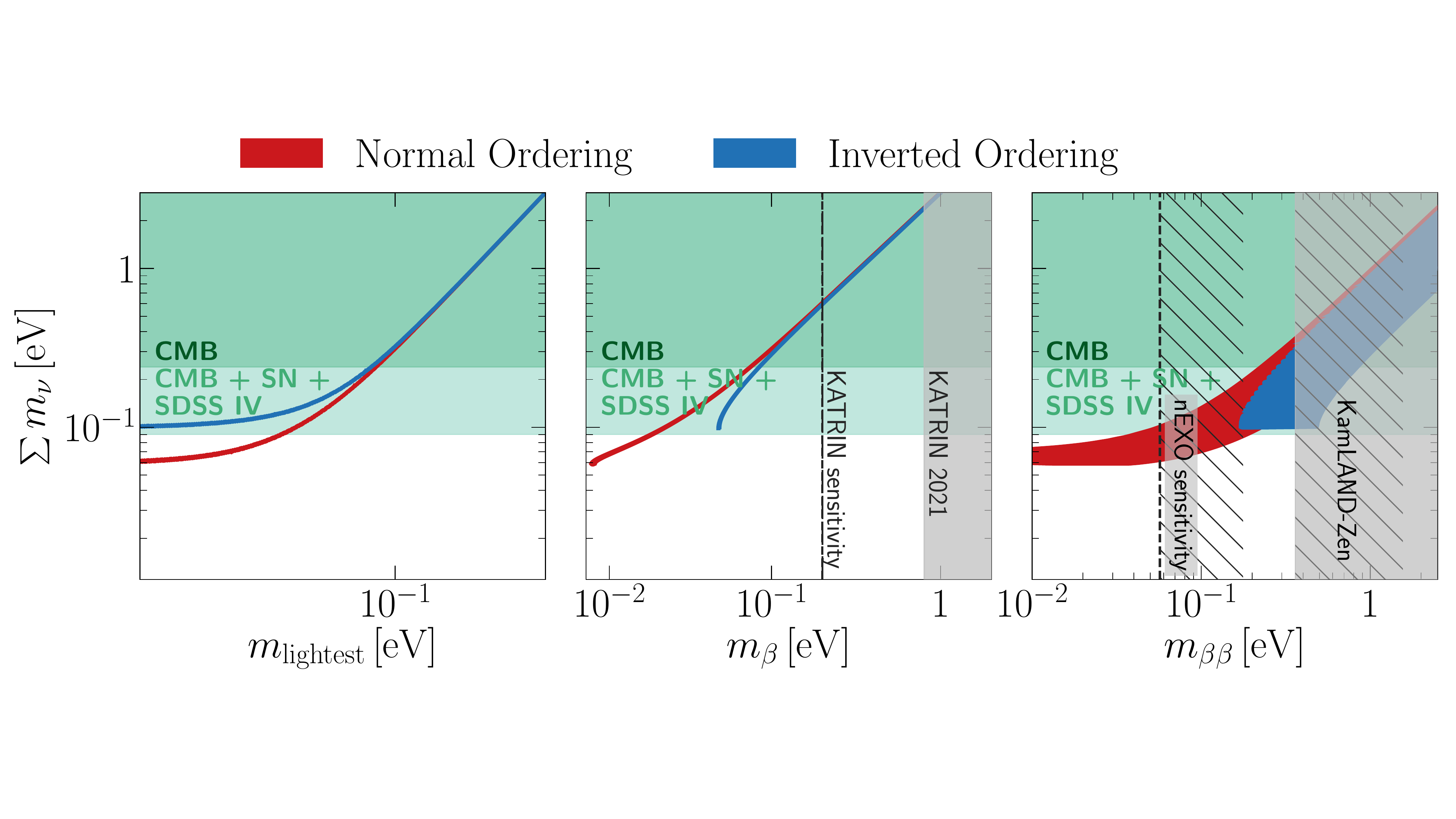}
	\caption{\textit{Theoretical predictions and current  bounds for the sum of neutrino masses $\sum m_\nu$ as a function of three quantities characterizing the neutrino masses: the lightest neutrino mass $m_{\rm lightest}$, beta-decay ($m_\beta$) and neutrinoless double beta decay ($m_{\beta\beta}$) effective masses are depicted in the left, middle and right panels respectively. The hatched regions in the right panel reflect the effect of uncertainties in the nuclear matrix elements in the bounds from neutrinoless double beta decay searches. Figure adapted from~\cite{Esteban:2020cvm}; see the reference for further information.}}
 \label{fig:f1}
\end{figure*}

\section{Cosmological bounds on neutrino masses}
\label{sec:mnu_bounds}

Relic neutrinos with sub-eV masses represent a good fraction (if not all) of the hot dark matter component in our current universe. These hot thermal relics leave clear signatures in the cosmological observables, see e.g.,~\cite{Lesgourgues:2018ncw, Lesgourgues:2006nd, Lattanzi:2017ubx, deSalas:2018bym, Vagnozzi:2019utt}, which can be exploited to put constraints on neutrino properties.

On the other hand, neutrino oscillations measured at terrestrial experiments indicate that at least two massive neutrinos exist in nature. Experiments measure two squared mass differences, the atmospheric $|\Delta m^2_{31}|=|m^2_3-m^2_1|= \approx 2.55\cdot 10^{-3}$~eV$^2$ and the solar $\Delta m^2_{21}=m^2_2-m^2_1 \approx 7.5\cdot 10^{-5}$~eV$^2$ splittings~\cite{deSalas:2020pgw, Esteban:2020cvm}. Since the sign of $|\Delta m^2_{31}|$ is unknown, two possible mass orderings are possible, the \emph{normal} ($\Delta m^2_{31}>0$) and the \emph{inverted} ($\Delta m^2_{31}<0$) orderings. In the normal ordering, $\sum m_\nu \gtrsim 0.06$~eV, while in the inverted ordering, $\sum m_\nu \gtrsim 0.10 $~eV.
As we shall see, current cosmological limits are approaching to the minimum sum of the neutrino masses allowed in the inverted hierarchical scenario, see Fig.~\ref{fig:f1}.
Cosmology can therefore help in extracting the neutrino mass hierarchy~\cite{Gariazzo:2018pei, RoyChoudhury:2019hls, Hannestad:2016fog, Lattanzi:2020iik}, which is a crucial ingredient in future searches of neutrinoless double beta decay~\cite{Agostini:2017jim, Giuliani:2019uno}. In the following sections, we shall review the main effects of neutrino masses in the different cosmological observables and the current bounds.

\subsection{CMB temperature, polarization and lensing bounds}
Traditionally, the main effects on the CMB of neutrino masses are those imprinted via the early ISW effect (previously described) and also that induced via changes in the angular location of the acoustic peaks, similarly to that discussed above for the case of $\Neff$. Concerning the ISW effect, 
the transition from the relativistic to the non-relativistic neutrino regime will get imprinted in the gravitational potential decays. As we have already explained, a larger (smaller) matter component will result into a smaller (larger) ISW amplitude. Therefore, an increase in the neutrino mass will induce a decrease in the height of the first CMB acoustic peak. There will also be a shift in the angular location of the acoustic peaks, that will move towards lower multipoles as we increase the value of the neutrino masses. 
Nevertheless, as in the case of $\Neff$, these two effects are strongly degenerate with other cosmological parameters such as the Hubble constant $H_0$, in such a way that a larger neutrino mass can always be compensated with a lower value of $H_0$.
But of course, we know that the CMB sets strong constraints on $\sum m_\nu$, how is this possible? The answer is via a secondary source of anisotropies, CMB lensing. The CMB photon path will be distorted by the presence of matter inhomogeneities along the line of sight between us and the last scattering surface. Lensing produces a remapping of the CMB anisotropies. The deflection angle induced by CMB lensing is proportional to the gradient of the lensing potential, which is sensitive to both the geometry of the universe and also to the matter clustering properties via the power spectrum of matter fluctuations. 
Neutrinos with sub-eV masses are hot thermal relics with very large velocity dispersions, and therefore they reduce clustering at scales smaller than their free streaming scale:
\begin{equation}
K_{\rm{fs,i}} \simeq\frac{0.677}{(1+z)^{1/2}}\left(\frac{m_{\nu,\rm{i}}}{1 \ \rm{eV}}\right)\Omega_{\rm m} ^{1/2} h \ \rm{Mpc}^{-1}~,
\end{equation}
where $\Omega_{\rm m}$ refers to the total matter component in the universe.
If we replace a massless neutrino component with a massive neutrino, the expansion rate increases and the growth of structure is suppressed. 
The net suppression of the power spectrum is scale-dependent and the relevant length scale is the Jeans length for
neutrinos, which decreases with time as the neutrino thermal velocities decrease.
Of course, this suppression of growth does not happen at scales larger than the neutrino free streaming scale. The net result is no effect on large scales and a suppression of power on small scales reducing, consequently, the lensing power spectrum~\cite{Kaplinghat:2003bh, Lesgourgues:2005yv}. Therefore, CMB lensing helps enormously in constraining neutrino masses and the Planck collaboration sets a bound $\sum m_\nu <0.24$~eV at $95\%$~CL from measurements of temperature, polarization, and lensing of the CMB~\cite{Aghanim:2018eyx} within the minimal $\Lambda$CDM model.
For neutrinos with degenerate masses, this implies that \emph{six million neutrinos cannot weigh more than one electron}.
Albeit current neutrino mass limits are very stable against extensions to the minimal $\Lambda$CDM cosmology (as we shall shortly  see), since CMB bounds mostly rely on lensing effects, the neutrino mass exhibits a non-negligible degeneracy with the lensing amplitude $A_{\rm L}$~\cite{Capozzi:2017ipn}.\footnote{The amount of lensing is a precise prediction of the $\Lambda$CDM model: the consistency of the model can be checked by artificially increasing lensing by a higher amplitude factor $A_{\rm{L}}$~\cite{Calabrese:2008rt} (\emph{a priori}, an unphysical parameter). If $\Lambda$CDM consistently describes all CMB data, observations should prefer $A_{\rm{L}}=1$. The so-called lensing anomaly~\cite{Planck:2018vyg,Motloch:2018pjy} is due to the fact that Planck CMB temperature and polarization observations prefer $A_\mathrm{L} > 1$ at $\sim 3\sigma$.}

\subsection{Large scale structure constraints}
Despite having non-negligible signatures in the CMB, it is precisely in large scale structure where the free streaming nature of neutrinos plays a major role. The information contained in the matter clustering in the universe can be interpreted in terms of measurements of the full-shape galaxy power spectrum or in terms of the Baryon Acoustic Oscillation (BAO) signal. A devoted study has found that the constraining power of the BAO signal is more powerful than that of the extracted power spectrum~\cite{Vagnozzi:2017ovm}, as it is less subject to e.g.\ non-linearities. Therefore, for constraining the neutrino mass, the BAO and the Redshift Space Distortions (RSD) are the usually exploited large scale structure observables. In the context of spectroscopic observations, the BAO signature can be exploited in two possible ways. Along the line of sight direction, BAO data provide a redshift dependent measurement of the Hubble parameter $H(z)$. Instead, across the line of sight, BAO data at the redshift of interest can be translated into a measurement of the angular diameter distance, which is an integrated quantity of the expansion rate of the universe $H(z)$.
In addition, anisotropic clustering in spectroscopic BAO measurements can also be exploited to extract RSD~\cite{Kaiser:1987qv}.
This effect, due to galaxy peculiar velocities, modifies the galaxy power spectrum and allows for an extraction of the product of the growth rate of structure ($f$) times the clustering amplitude of the matter power spectrum ($\sigma_8$), the well-known $f\sigma_8$ observable.
Analyzing current cosmological data from the Planck CMB  satellite, the SDSS-III and SDSS-IV galaxy clustering surveys~\cite{Dawson:2015wdb,Alam:2020sor} and the Pantheon Supernova Ia sample, Ref.~\cite{DiValentino:2021hoh}
found one of the most constraining neutrino mass bounds to date, $\sum m_\nu<0.09$~eV at $95\%$~CL, mostly due to RSD analyses from the SDSS-IV eBOSS survey (see also Ref.~\cite{Palanque-Delabrouille:2019iyz} for a similar limit). 
Such constraint on the sum of neutrino masses, when one considers neutrino oscillation constraints on the mass differences, implies that \emph{one electron is heavier than at least nine and a half million of the heaviest neutrinos}.\footnote{Given the limit $\sum m_\nu<0.09$~eV and $\Delta m^2_{31}\sim2.5\cdot10^{-3}$~eV$^2$, the heaviest neutrino has a mass $m_{\rm heaviest}\lesssim53$~meV, while the lightest one corresponds to $m_{\rm lightest}\lesssim17$~meV. Notice that based on the value of $\Delta m^2_{31}$ alone, the heaviest neutrino cannot have a mass smaller than $\sim50$~meV, obtained when the lightest neutrino is massless.}
We depict these limits in Fig.~\ref{fig:f1}, together with present and future sensitivities from beta decay laboratory experiments and searches from neutrinoless double beta decay probes.
More recently, the DESI collaboration provided new observations of the BAO scale,
which strengthen the limit on the sum of the neutrino masses to $\sum m_\nu<0.072$~eV (95\% CL) when combined with CMB data~\cite{DESI:2024mwx}.

Firstly, we would like to remark that these neutrino limits are extremely robust and solid, and therefore very stable against fiducial cosmologies. Table~\ref{tab:tab2} depicts the limits on $\sum m_\nu$ in a number of different underlying cosmologies. Notice that the limits obtained within the minimal $\Lambda$CDM model remain almost unmodified except for some particular models, such as when the lensing amplitude is a free parameter, due to the large impact of neutrino masses on CMB lensing. The model-marginalized limit obtained in Ref.~\cite{diValentino:2022njd} is $0.1$ eV, extremely tight and very close to the expectations within the inverted ordering for the minimum value of the sum of the neutrino masses. Current neutrino mass limits are therefore very difficult to avoid within the $\Lambda$CDM framework and its extensions, and to relax them one would need to search for non-standard neutrino physics, such as exotic beyond Standard Model interactions or decays, and/or modified gravitational sectors.

Secondly, as the upper bound on the neutrino mass approaches the minimal prediction within the inverted ordering scenario, one might claim the rejection of the former at a given significance level. Plenty of debate and various studies in the literature have been devoted to settling this issue, see e.g.~\cite{DeSalas:2018rby,Gariazzo:2018pei,Simpson:2017qvj,Jimenez:2022dkn,Gariazzo:2022ahe,Gariazzo:2023joe}. Recently, in~\cite{Gariazzo:2023joe}, the authors quantified the current preference for the normal mass ordering versus the inverted one using the Bayes factor.
None of the cases explored by the authors (i.e., using terrestrial data alone or current cosmology without terrestrial data) show a particularly significant preference for the normal mass ordering. The same reference indicates that future cosmological experiments, expected to achieve a $1\sigma$ precision on $\sum m_\nu$ at the level of 0.02 eV, will not provide a strong preference in favor of the normal ordering (if nature has chosen this scenario), reaching a $2-3\sigma$ significance at most.

\begin{table}
\centering
\begin{tabular}{ccc}
\toprule
\textbf{Cosmological model} & & \boldmath{$\sum m_\nu$\textbf{[eV]}}\\
\hline\hline

$+\sum m_\nu$
&  DH & $<0.0866$\\
&  NH & $<0.129$\\
&  IH  & $<0.155$\\
[2ex]

$+\sum m_\nu+\Neff$
&  DH & $<0.0968$ \\
&  NH & $<0.131$ \\
&  IH & $<0.163$ \\[2ex]

$+\sum m_\nu+\Omega_{k}$
&  DH & $<0.111$\\
&  NH & $<0.143$\\
&  IH & $<0.180$\\[2ex]

$+\sum m_\nu+\alpha_s$
&  DH & $<0.0908$\\
&  NH & $<0.128$\\
&  IH &$<0.157$ \\[2ex]

$+\sum m_\nu+r$
&  DH & $<0.0898$\\
&  NH & $<0.130$\\
&  IH & $<0.156$\\[2ex]

$+\sum m_\nu+w_0$
&  DH & $<0.139$\\
&  NH & $<0.165$\\
&  IH & $<0.204$\\[2ex]

$+\sum m_\nu+(w_0>-1)$
&  DH &$<0.0848$\\
&  NH & $<0.125$\\
&  IH & $<0.157$\\[2ex]

$+\sum m_\nu+w_0+w_a$
&  DH & $<0.224$\\
&  NH & $<0.248$\\
&  IH & $<0.265$\\[2ex]

$+\sum m_\nu+A_{\text{L}}$
&  DH & $<0.166$\\
&  NH &$<0.189$\\
&  IH & $<0.216$\\[2ex]

\hline
model marginalized & DH & $<0.102$ \\

\bottomrule
    \end{tabular}
    \caption{Constraints at 68\% and upper limits at 95\% CL,
    for the $\Lambda$CDM+$\sum m_\nu$ model and its extensions (adapted from Ref.~\cite{diValentino:2022njd}).}
\label{tab:tab2}
\end{table}

\section{Forecasts for Future experiments}
\label{future}

In the near future, constraints on the total neutrino mass and the number of neutrino species will significantly benefit from data from the CMB and Large Scale Structure observations.

Ground-based CMB telescopes~\cite{Chang:2022tzj} currently represent the proposals with the highest likelihood of realization. However, they must be complemented by measurements at large angular scales (such as those from Planck or future experiments) and a thorough understanding of foregrounds to effectively narrow the uncertainties in the neutrino sector.

The Simons Observatory (SO)~\cite{SimonsObservatory:2019qwx} will be the first experiment poised to enhance neutrino constraints. It aims to measure the total neutrino mass with an uncertainty of $\sigma (\sum m_\nu) = 0.04$~eV when combined with DESI BAO data~\cite{DESI:2016fyo,DESI:2024mwx} and weak lensing data from the Rubin Observatory~\cite{lsstsciencecollaboration2009lsst}.
By leveraging LiteBIRD's~\cite{LiteBIRD:2020khw} forthcoming cosmic variance-limited measurements of the optical depth to reionization $\tau$, SO can achieve a precision of $\sigma (\sum m_\nu) = 0.02$~eV for the total neutrino mass.
Furthermore, SO aims to determine the number of neutrino species, $N_{\rm eff}$, with an accuracy of $\sigma (N_{\rm eff}) = 0.07$.

Next-generation Stage IV CMB experiments, such as CMB-S4~\cite{Abazajian:2019eic}, are aiming to determine \Neff\ with an uncertainty of $\leq 0.06$ at the 95\% CL.
When combined with BAO measurements from DESI and the current Planck measurement of the optical depth, CMB-S4's observations of the lensing power spectrum and cluster abundances will yield a constraint on the sum of neutrino masses of $\sigma (\sum m_\nu) = 0.024$~eV. This constraint would improve to $\sigma (\sum m_\nu) = 0.014$~eV with more accurate future measurements of the optical depth.

Alternatively, upcoming proposals for future CMB telescopes, such as PICO~\cite{NASAPICO:2019thw}, when combined with BAO data from DESI or Euclid~\cite{EUCLID:2011zbd}, are expected to achieve an accuracy of $\sigma (\sum m_\nu) = 14$~meV. This corresponds to a $4\sigma$ detection of the minimum total neutrino mass, as anticipated for the NH.
Additionally, these telescopes should constrain $\Delta N_{\rm eff} < 0.06$ at the 95\% CL. A CMB telescope stands out as the sole instrument capable of precisely measuring all these neutrino properties, along with the optical depth, using a single dataset. This approach avoids the challenges associated with cross-calibration.

Finally, CMB-HD~\cite{CMB-HD:2022bsz} represents a futuristic millimeter-wave survey that is anticipated to achieve an uncertainty on \Neff\ of $\sim 0.014$ at the 68\% CL and $\sigma (\sum m_\nu) = 0.013$~eV (a minimum $5\sigma$ detection for the sum of neutrino masses). This precision will be attained through measurements of the gravitational lensing of the CMB and the thermal and kinetic Sunyaev-Zel'dovich (SZ) effects on small scales.

\section{Direct detection of the relic neutrino background}
\label{sec:direct_detection}
As we discussed in the previous sections,
we have rather clear indirect evidence of the existence of the \cnb\ due
to cosmological observables constraining \Neff\ to be very close to the expected value \Neffstd.
However, these results
cannot definitively tell us that the amount of radiation energy density observed in the universe is truly associated with standard model neutrinos.
Such confirmation would only come from a direct detection of the \cnb,
which would demonstrate the presence of neutrinos with the expected momentum distribution and temperature.

As we mentioned in Section~\ref{sec:nu_decoupling},
relic neutrinos are predicted to have a momentum distribution function that is a slightly distorted Fermi-Dirac distribution,
with a temperature of approximately $0.1$~meV.
Consequently, the average neutrino energy would be $\sim0.5$~meV.
Neutrino interaction cross sections, however,
decrease rapidly with their energy, making the detection of such relic neutrinos an extremely challenging task.
In the past, several authors studied the problem of direct detection of the \cnb\ and proposed various experimental methods which can, in principle, allow us to observe a signal from relic neutrinos.
A detailed review of all the methods can be found in~\cite{Bauer:2022lri},
and here we summarize the main proposals.

\subsection{Neutrino capture on beta-decaying nuclei}
In 1962, Weinberg~\cite{Weinberg:1962zza} proposed a method to detect the presence of ``a shallow degenerate Fermi sea of neutrinos'' that fills the universe.
During those times, neutrinos were believed to be massless,
and the proposed method required using a beta-decaying nucleus,
which could decay and emit an electron if capturing a neutrino from the \cnb:
the process is therefore called ``neutrino capture on beta-decaying nucleus''.
The original proposal considered a possible depletion in the electron (positron) energy spectrum
due to a large chemical potential of the neutrino.
The process has no energy threshold, so it is not a problem if neutrinos have very small energy.
In 2007, the authors of~\cite{Cocco:2007za} revisited the original proposal
to properly describe the effect of neutrino capture in the absence of large chemical potentials
but in the presence of neutrino masses,
and discussed which beta-decaying nuclei can best serve the purpose.
One of the crucial points is that the neutrino capture process is related to standard beta-decay,
with the difference that the neutrino (or antineutrino) is in the final state
and the energy of the electron can exceed the end-point value $E_0$ of the beta-decay spectrum.
Therefore, in order to build a successful experiment,
it is crucial to be able to distinguish neutrino capture events with energy above $E_0$
from beta-decay events with energy below $E_0$.
This is no easy task, because the energy separation between beta-decay and neutrino capture events
is equal to twice the neutrino mass, see e.g.~\cite{Long:2014zva},
and the amount of background versus signal events is huge.\footnote{If the energy resolution is not sufficiently good,
it has been shown that a direct detection of the \cnb\ is still possible if one can detect the periodicity of the signal,
generated by the peculiar motion of the laboratory in the \cnb\ rest frame~\cite{Akhmedov:2019oxm},
over the beta-decay background.
However, the required number of observed events is still far from any proposed realization of the experiment. }
The authors of~\cite{Cocco:2007za} consider the half-life and cross section of different nuclei
and determine that the best chances to build an experiment emerge when adopting tritium:
it provides a reasonably large event rate for neutrino capture together with a
sufficiently small contamination of the signal region by beta-decay background events.

Based on this result, the first experimental attempt at detecting the \cnb\ by neutrino capture
on tritium is being developed at Gran Sasso Laboratories in Italy.
The PTOLEMY proposal~\cite{Betts:2013uya,PTOLEMY:2018jst} plans to reach a final setup
with approximately 100~g of tritium and a final energy resolution in the ballpark of 0.1eV
in order to detect $\sim5$ neutrino capture events per year if their separation from the
beta-decay spectrum is sufficiently large.
According to the first set of simulations, this setup could guarantee a $3\sigma$ observation of the \cnb\ in one year
if neutrino masses are above 0.2eV~\cite{PTOLEMY:2019hkd}.
Even if this is not true for standard neutrinos,
PTOLEMY could still detect the presence of sterile neutrinos in the \cnb.
In the case of sufficiently good energy resolution, the number of signal events could be enhanced by
a larger local number density of relic neutrinos~\cite{deSalas:2017wtt},
non-standard neutrino interactions~\cite{Arteaga:2017zxg,Banerjee:2023lrk},
or if neutrinos are Majorana particles~\cite{Roulet:2018fyh,Hernandez-Molinero:2022zoo,Perez-Gonzalez:2023llw}.

\subsection{Elastic scattering on macroscopic targets}
Another way to detect relic neutrinos makes use of their elastic scattering on macroscopic targets.
Under this category, we can find two separate effects.
The first one was proposed by Stodolsky~\cite{Stodolsky:1974aq} and revisited by~\cite{Duda:2001hd}.
In the presence of a background of relic neutrinos,
the two spin states of the electron are modified by the presence of either an asymmetry between neutrinos and antineutrinos~\cite{Stodolsky:1974aq},
or if there is a net helicity asymmetry in the \cnb~\cite{Duda:2001hd}.
It is important to notice that the Stodolsky effect is proportional to the Fermi constant $G_F$,
while normally neutrino cross-sections are suppressed by $G_F^2$.
The energy split of the spin states generates a torque on electrons.
In the case of a ferromagnet, the effect will generate a torsion that, in principle, can be measured
with a torsion balance.
Although Cavendish-style torsion balances are not sensitive enough to measure the extremely small torque that could be associated with the \cnb,
torsion balances where the test masses are suspended by superconducting magnets
could provide much better perspectives, see e.g.~\cite{Bauer:2022lri} and references therein.

If instead we consider the fact that the de Broglie length $\lambda_\nu$ of relic neutrinos is very large due to their tiny momentum,
we can have a strong enhancement of neutral current scattering of relic neutrinos
on the nuclei in the test mass, see~\cite{Shergold:2021evs,Domcke:2017aqj}.
It can be shown that for relic neutrinos,
the enhancement of the cross-section is proportional to the number of target nuclei
within a volume with a side equal to the de Broglie length,
which has macroscopic values of $\mathcal{O}({\rm mm})$ given their tiny energy.
As a consequence, the coherent scattering cross-section is significantly larger than its microscopic (scattering off single nuclei) coherent counterpart.
In addition, neutrinos can also coherently scatter off electrons in the target mass:
in this case, too, the cross-section is significantly enhanced, but the recoil of electrons
may not be transferred completely to the target atoms.
Despite the coherence factor, the acceleration generated by relic neutrinos on the test mass
is extremely small.
To detect such accelerations,
it has been proposed to measure the tiny strains on laser interferometers used to detect gravitational waves~\cite{Domcke:2017aqj,Lee:2020dcd}
because they have much more precision at detecting small variations of distances.

\subsection{Neutrino capture on accelerated ion beams}
We discussed how we must consider thresholdless processes to detect relic neutrinos.
An interesting idea considered in~\cite{Bauer:2021txb} makes use of accelerated ion beams.
By colliding the ions with \cnb\ particles,
one can meet the threshold required for certain neutrino capture processes
in the center-of-mass frame of the system
and avoid the thresholdless requirement.
In this way, one can also tune the neutrino energy to hit a resonance and enhance the neutrino capture cross-section.
Once the ionized beam hits the \cnb,
the neutrino capture process converts the original ion into an unstable one.
As a consequence,
it may be difficult to estimate the performance of the experiment by the neutrino capture rate,
because the presence of the unstable daughter states decreases over time,
and the conversion rate from the original state to the decaying state quickly reaches a maximum
if the decay rate equals the neutrino capture rate.
For this reason, a better idea involves nuclei that can undergo a 3-state resonant bound beta decay:
\begin{eqnarray}
{}^A_ZP+\nu_e
&\rightarrow&
{}^A_{Z+1}D+e^-\mbox{ (bound)}
\nonumber\\
&\rightarrow&
{}^A_{Z+2}F+2e^-\mbox{ (bound)}+\bar\nu_e
\,,
\end{eqnarray}
where $A$ and $Z$ are the mass and atomic number of the parent state $P$,
while $D$ and $F$ are the daughter and final states, respectively.
If $F$ is a stable state different from $D$ and $P$, after $D$ decays, it remains indefinitely
in the $F$ state, which can be easily measured.
To maximize the detection perspectives, it can be shown~\cite{Bauer:2021txb,Bauer:2022lri}
that processes with low $Q$ values are preferred.
Even with $Q$ values on the order of tens to a few hundred keV, beam energies
of hundreds to several thousand TeV are required.
The perspectives can be improved by considering excited states in the beam,
which reduce the required energy threshold,
but at the expense of beam stability and increased experimental challenges.

\section{Summary and conclusions}
Neutrino physics plays an important role in the evolution of the universe.
Neutrinos influence the late phases of radiation domination and govern the amount of light element abundances obtained from BBN.
As relativistic particles, their presence affects the matter-radiation equality epoch and the formation of the CMB.
When they become non-relativistic particles, their free-streaming properties impact the growth of structures in the late universe.
Here, we discussed the contribution of neutrinos to all these processes,
starting from how neutrinos decouple from the thermal plasma and what their momentum distribution function is,
and then detailing how cosmological observables can help us to learn more about these elusive particles.
For instance, current cosmological measurements
confirm that there are approximately three neutrino-like relativistic particles
in the early universe,
providing us with an indirect probe of their existence.
After their decoupling during radiation domination,
neutrinos redshift and their temperature decreases,
until eventually, most of them become non-relativistic at some point during matter domination.
Following the non-relativistic transition, neutrino free-streaming imprints
a characteristic signature on the matter power spectrum,
allowing us to derive strong bounds on their total energy density, largely arising from large-scale structures.
Such bounds can be converted into limits on the sum of the neutrino masses,
assuming that neutrinos are stable and their mass remains constant over the lifetime of the universe.
Under such conditions, the cosmological bound on neutrino masses
is much stronger than the limits obtained by terrestrial experiments nowadays.
Cosmological probes, however, seem to prefer a null value for the sum of the neutrino masses,
possibly in tension with requirements imposed by the existence of neutrino oscillations.
We also comment on the capabilities of incoming and future cosmological observation in constraining the amount of neutrinos in the universe.
Non-relativistic neutrinos also feel the gravitational attraction of matter structures and may cluster at small scales,
so their number density is not constant throughout the entire universe.
Instead, there are overdensities where large distributions of dark matter and baryons are located, such as within galaxies or clusters of galaxies,
including at our position in the Milky Way.
Finally, we discuss how the relic neutrinos from the early universe could be detected
in terrestrial experiments.
The direct detection of relic neutrinos would firmly confirm that the relativistic degrees of freedom we observe through CMB and BBN observables are truly the standard model neutrinos,
but the extreme difficulty of direct detection experiments
(especially due to the tiny energy distribution relic neutrinos have and the consequently feeble cross section)
makes this goal still quite distant.
Efforts in developing a suitable direct detection experiment, however, are ongoing, such as the PTOLEMY proposal.
A direct detection of relic neutrinos would revolutionize our understanding of the early universe in several ways:
we would be certain about the existence and stability of neutrinos across the entire universe history, and
we would learn if deviations from the standard cosmological model occurred at epochs that the CMB cannot test.
We might also be able to study whether neutrinos behave in the expected way in the gravitational potential of local structures and how they cluster in the local neighborhood. Cosmological neutrinos are therefore critical to establishing a robust link not only between the predictions from both standard and non-standard neutrino scenarios and particle physics but also between the canonical growth of structure and the different models for the large scale structures we observe today in our universe.

\bibliographystyle{apsrev4-1}
\bibliography{main}

\end{document}